\documentclass[conference]{IEEEtran}
\IEEEoverridecommandlockouts
\usepackage{cite}
\usepackage{amsmath,amssymb,amsfonts}
\usepackage{algorithmic}
\usepackage{graphicx}
\usepackage{textcomp}
\usepackage{xcolor}
\usepackage{hyperref}
\usepackage{booktabs}
\def\BibTeX{{\rm B\kern-.05em{\sc i\kern-.025em b}\kern-.08em
    T\kern-.1667em\lower.7ex\hbox{E}\kern-.125emX}}
\begin{document}

\title{PEFSL: A deployment Pipeline for Embedded Few-Shot Learning on a FPGA SoC}


\author{\IEEEauthorblockN{
Lucas Grativol\IEEEauthorrefmark{1}\IEEEauthorrefmark{3},
Lubin Gauthier\IEEEauthorrefmark{1},
Mathieu Léonardon\IEEEauthorrefmark{1},
Jérémy Morlier\IEEEauthorrefmark{1},
Antoine Lavrard-Meyer\IEEEauthorrefmark{1},\\
Guillaume Muller\IEEEauthorrefmark{3},
Virginie Fresse\IEEEauthorrefmark{2} and
Matthieu Arzel\IEEEauthorrefmark{1}
}
\IEEEauthorblockA{\IEEEauthorrefmark{1}IMT Atlantique, Lab-STICC, UMR CNRS 6285, F-29238 Brest, France}
\IEEEauthorblockA{\IEEEauthorrefmark{2}Hubert Curien Laboratory, Saint-Etienne, France}
\IEEEauthorblockA{\IEEEauthorrefmark{3}Mines Saint-Etienne, Institut Henri Fayol, Saint-Etienne, France}}

\maketitle

\begin{abstract}
This paper tackles the challenges of implementing few-shot learning on embedded systems, specifically FPGA SoCs, a vital approach for adapting to diverse classification tasks, especially when the costs of data acquisition or labeling prove to be prohibitively high. Our contributions encompass the development of an end-to-end open-source pipeline for a few-shot learning platform for object classification on a FPGA SoCs. The pipeline is built on top of the Tensil open-source framework, facilitating the design, training, evaluation, and deployment of DNN backbones tailored for few-shot learning. Additionally, we showcase our work's potential by building and deploying a low-power, low-latency demonstrator trained on the MiniImageNet dataset with a dataflow architecture. The proposed system has a latency of 30 ms while consuming 6.2 W on the PYNQ-Z1 board.
\end{abstract}

\section{Introduction}
For object classification, the conventional approach involves training a neural network using a big labeled dataset. However, such datasets are not always available, usually because the cost of labeling is high~\cite{fredriksson2020data}. Another, more innovative method is to use a pre-trained network and to specialize it on few labeled examples. This can be performed with transfer learning or fine-tuning~\cite{weiss2016survey}. But when the number of labeled examples is really low, the method to be used is called \emph{few-shot learning}~\cite{b2}. Few-shot learning seeks to leverage the knowledge from deep learning (DL) models to achieve robust classification performance on new tasks, when only a handful of labeled samples per class are available. 

One of the primary obstacles to the implementation of few-shot learning on embedded systems is the required computational power induced by the underlying cost of Deep Neural Networks (DNN) models. Careful design of low-complexity DNN adapted to embedded hardware targets is therefore a main concern~\cite{ahmad2020accelerating}. Among the potential hardware that can be found in embedded systems, FPGA SoCs (System-On-Chip) have proven to be good candidates for the deployment of DNNs when energy consumption is critical~\cite{b10,b9} or when low latency is at stake~\cite{zhang2021low}. However, there have been few examples of such FPGA implementations in the literature so far. The challenges to be tackled toward such an implementation are the selection and adaptation of deployment frameworks, the identification and adaptation of an efficient training routine from the literature, and finally the design of a lightweight network that meets the constraints of embedded systems while also performing well for the defined task, few-shot learning on embedded FPGA SoC, with a real-time classification of a video stream.\\
In this paper, we tackle these challenges. Our contributions can be summed up as:

\begin{itemize}
    \item one of the first few-shot learning platforms for real-time object classification on an FPGA SoC in the literature,
    \item a full open source pipeline\footnote{\url{https://github.com/brain-bzh/PEFSL}}, based on the Tensil framework\footnote{\url{https://www.tensil.ai/}}, for the design, training, evaluation and deployment of DNN backbones for few-shot learning on FPGA SoCs,
    \item the demonstration of the potential of this platform on a use case, the design and deployment of a low power and low latency few-shot model on the MiniImageNet dataset on a given hardware architecture.
\end{itemize}

These contributions aim to pave the way for exciting new applications in fields such as robotics, drones, and autonomous vehicles, where responsiveness, computational power, and energy are critical factors. The entire source code needed to replicate all aspects of this work are open source.

\section{Few-Shot Learning}

\begin{figure}[t]
\flushright{\includegraphics[width=0.53\textwidth]{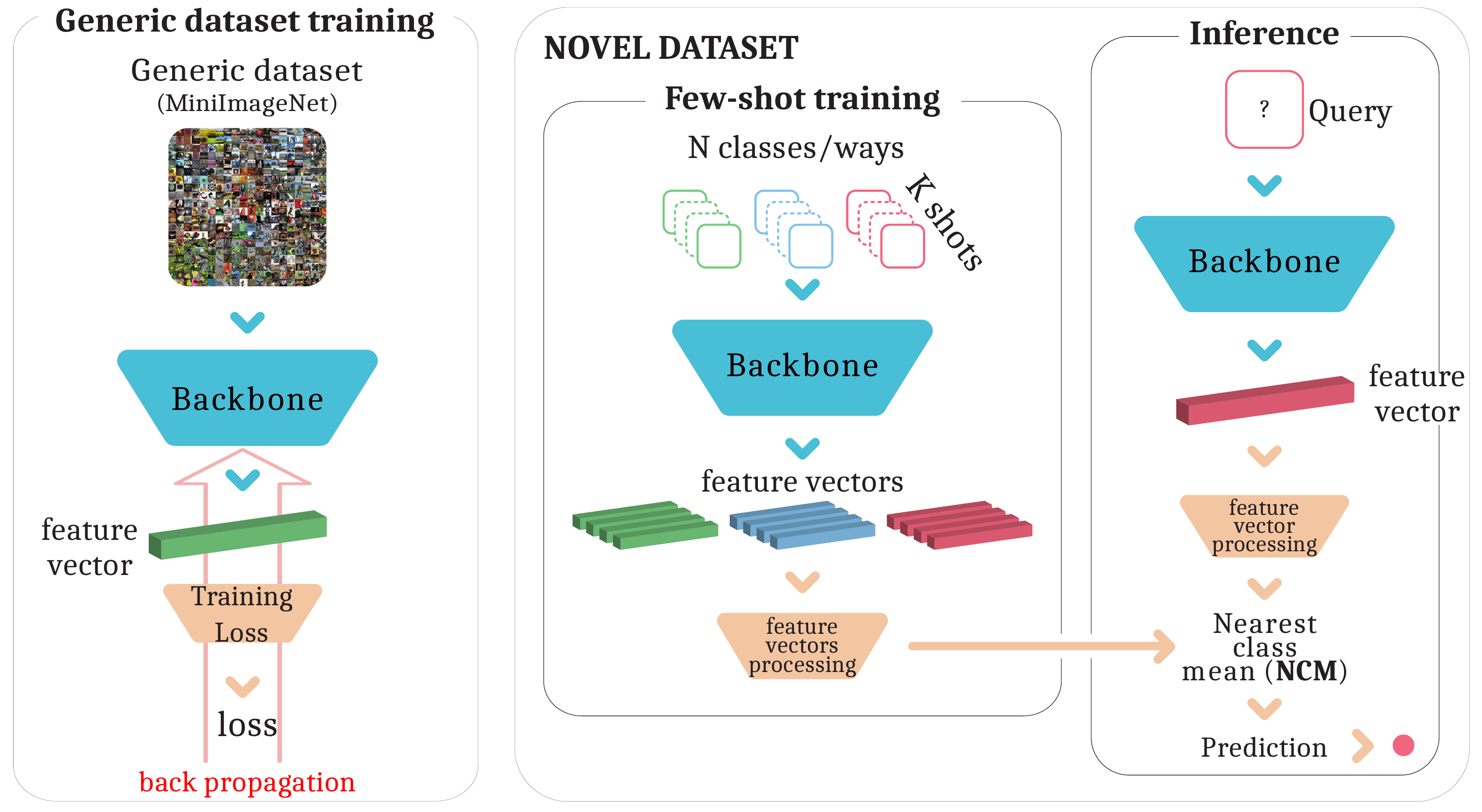}}
\caption{Our few-shot learning method.}
\label{few_shot_learning}
\end{figure}

Few-shot learning consists in classifying examples for unseen classes with a very small number of training examples. State-of-the-art methods are based on DL approaches. This may, at first, be counter-intuitive as DL is known to perform well when fed with huge databases on which it excels at generalizing.

Our few-shot paradigm is depicted in Fig.~\ref{few_shot_learning}. The first step (generic dataset training) consists in training a DNN, called backbone, following the method detailed in~\cite{b2}. This is performed by training a classification network with an additional pretext loss~\cite{gidaris2019boosting}. Few-shot datasets are usually split between the \emph{base} and \emph{validation} dataset, the latter being used to assess the generalization performance of the model. On the contrary to standard classification datasets, in the case of few-shot learning, the classes of the validation set are distinct from those of base set~\cite{mangla2020charting}, in order to evaluate the generalization performance on new classes. Once trained, the backbone is kept frozen for the subsequent steps, as its only function is to map the input to an high dimension, the feature vectors.

Then, in the next step, the few-shot learning performance is evaluated on a third set of images called the \emph{novel} dataset. This novel dataset consists of thousands of few-shot episodes~\cite{laenen2021episodes}. In each episode there is a certain number of classes, called \emph{ways}. For each way, there will be a given number of labeled examples called \emph{shots} and some unlabeled ones called \emph{queries}, as depicted in the "Few-shot training" and "Inference" diagrams in the Fig.~\ref{few_shot_learning}. The performance of the model corresponds to the number of queries that are correctly identified using the few available shots, averaged on the thousands of episodes. The number of shots and ways are set by the benchmarks. As an important distinction in the few-shot learning domain, we aim to solve an inductive~\cite{scott2018adapted} problem, when one doesn't have access to the whole set of queries beforehand, and not a transductive~\cite{liu2018learning} one, where one has access to the queries.

Because the DNNs used as backbones are usually complex in terms of memory footprint and computational complexity, the efficiency of these methods in embedded environments remains a challenge. Though, rapid adaptation to new tasks using minimal resources is essential, especially for applications such as real-time object recognition on embedded systems like drones or autonomous robots. Therefore, specific effort has to be made on the design of the backbones.

\section{Backbones}
\label{sec:backbones}

\subsection{Architecture}

In this experiment, we use ResNets, adapted from~\cite{he2016deep}. The primary feature of a ResNet is the use of residual blocks, where bypasses between certain layers of the network are added. The main advantage of this network architecture is the ability to train much deeper and more accurate networks than traditional Convolutional Neural Networks (CNN) such as VGGs or AlexNet~\cite{li2021survey}. This type of network is particularly efficient for our experiment. Indeed, they allow for performance that is very close to the state of the art~\cite{b2}. Though, they are small networks with relatively few parameters and limited computational complexity.

\begin{figure}[t]
\centerline{\includegraphics[width=0.47\textwidth]{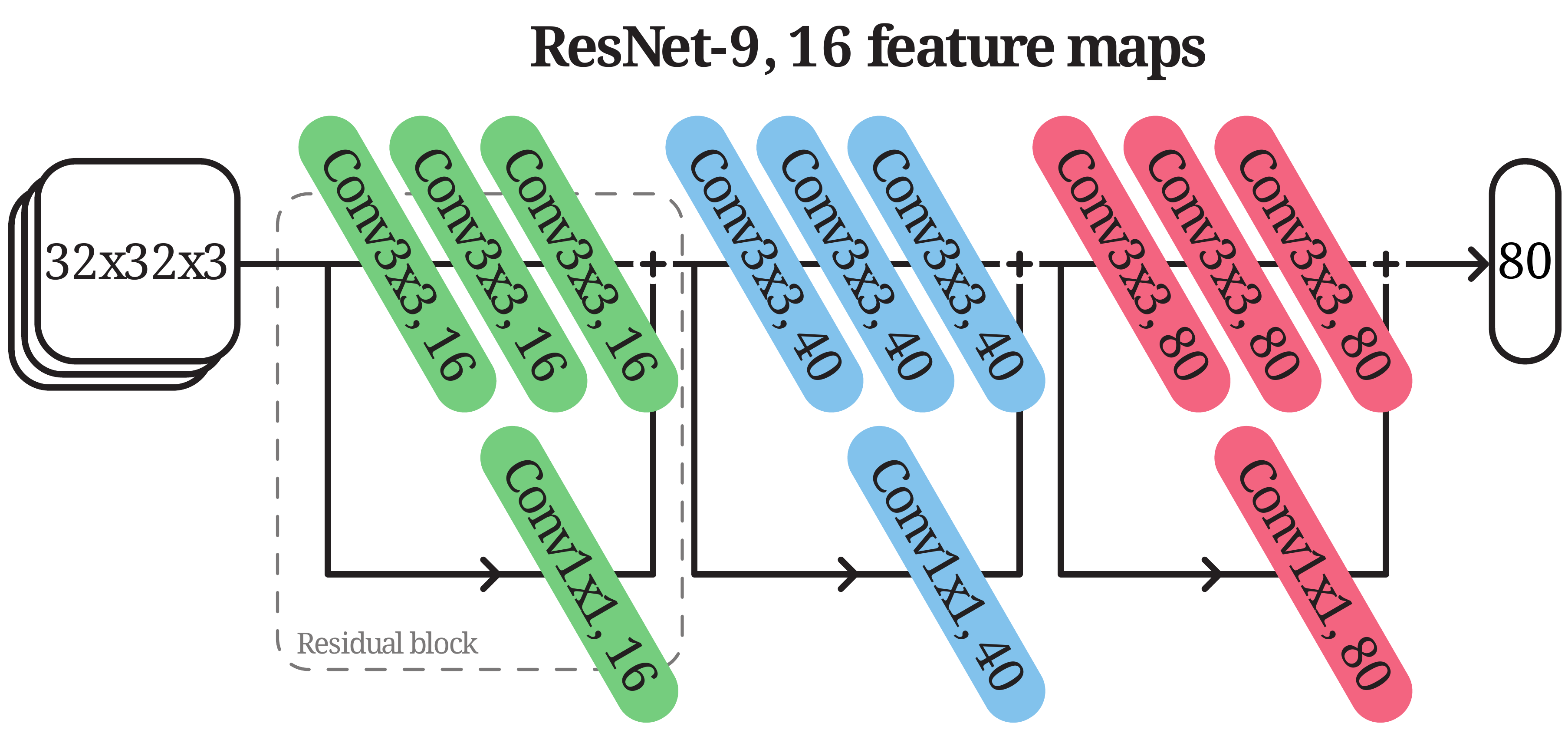}}
\caption{Structure of a ResNet-9, where initial layers employ 16 output feature maps, and subsequent layers scale their output channels accordingly.}
\label{resnet9}
\end{figure}


\subsection{Hyperparameters}

Here, we list the main hyperparameters that influence the final system performance and its complexity:

\paragraph{Network depth} We choose two ResNet architectures with small depths, ResNet-9 and ResNet-12. A ResNet-9 is simply a ResNet-12 with the last residual block removed. It is expected that the shallower and less computationally intensive ResNet-9 may exhibit lower accuracies for complex tasks. This specific ResNet-9 architecture is depicted in Fig.~\ref{resnet9}.
\paragraph{Training and test image size} The size of training images impacts the amount of computation to be performed, and it also affects performance. Smaller images, like $32\times32$, contain less information than $100\times100$ images, but processing them requires fewer operations. As we will see in section~\ref{sec:results}, the joint choice of testing and training image size resolution has a huge impact on the accuracy of the model.
\paragraph{Downsampling} Between each residual block, the resolution of the feature maps is reduced. We have two ways to perform this reduction. Either we change the strides of the last convolution in each block from 1 to 2, or we use max pooling, which consists in retaining only the maximum value of groups of values in the feature maps. A stride of 2 or a $2\times2$ pooling size are equivalent in terms of dimension reduction.
\paragraph{Number of feature maps} The backbone is mainly composed of convolution layers. Where the number of filters used on a layer defines the number of feature maps output by that layer. We set the number of filters in the first convolution layer as a hyperparameter, scaling subsequent layers accordingly.

\subsection{Training}


We use the MiniImageNet~\cite{vinyals2016matching} dataset, extracted from ImageNet~\cite{deng2009imagenet}. It consists of 64 base classes, 16 validation classes, and 20 novel classes. Each class contains 600 images, and the resolution is $84\times84$. In this paper, we focused on the 5-ways, 1-shot setup. Nevertheless, it has been noticed that performance of a given model and training routines are usually closely correlated across different numbers of shots~\cite{b2}. The MiniImageNet dataset is specifically designed for few-shot learning. Its value lies in the fact that it contains highly diverse classes that allow for excellent generalization to new classes.

\section{Open Source Pipeline }

\subsection{PEFSL pipeline}

\begin{figure}[t]
\centerline{\includegraphics[width=0.4\textwidth]{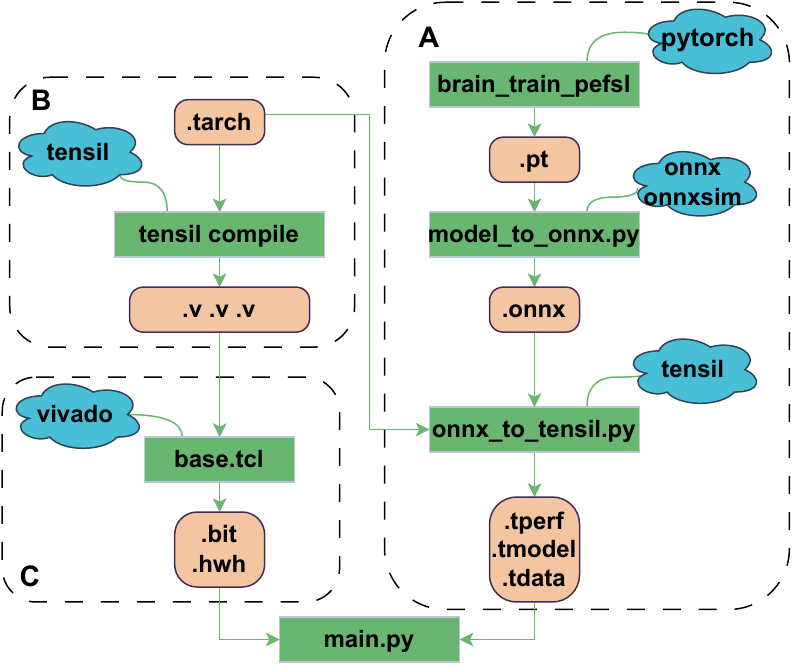}}
\caption{Modular pipeline for the deployment of a few-shot learning system on an FPGA SoC.}
\label{pipeline}
\end{figure}

In order to explore the search space of the previously defined training and network architectures hyperparameters, we developed and released PEFSL, a modular pipeline for the training, compilation, hardware synthesis and deployment of a few-shot learning application on an FPGA SoC. It uses several tools that will be detailed hereafter and are depicted in Fig.~\ref{pipeline}.
Part \textbf{A} of PEFSL corresponds to the training routine of the backbone, as described in section \ref{few_shot_learning}, its conversion in the ONNX format, and its compilation with the Tensil framework. Tensil is an open-source framework for running machine learning models on custom accelerator architectures. The training routine is adapted from~\cite{b2} in which we added the ResNet-9 and ResNet-12 architectures and their variants. It is using state-of-the-art techniques for training such CNNs on few-shot learning tasks. Then, the pytorch model is translated into an ONNX format. We also use the ONNX simplifier tool that allows for efficient simplification on the ONNX model. Finally, this ONNX model is compiled with the Tensil framework. Provided a description of the underlying architecture (\texttt{.tarch} file), that specifies the features of the systolic arrays~\cite{xu2023survey} (number of Processing Elements, data format, memory size). This first three scripts allow for generating automatically the latency of the neural network on the given architecture. Therefore it can be used to perform a design space exploration of the neural network architectures and training techniques, such as in Fig.~\ref{result_implem}.

Part \textbf{B} corresponds to the compilation of the architecture that generates RTL files of the Tensil accelerator IP. These RTL files are used in part \textbf{C}, which provides project files to the AMD-Xilinx Vivado tool that generates the bitstream of the PL (Programmable Logic) used in the demonstrator. The produced intermediary files (bitstream and Tensil model) are then used in the main script, which uses the PYNQ driver, that is used for the data transfer between the CPU and the FPGA.

\subsection{Demonstrator}

\begin{figure}[t]
\centerline{\includegraphics[width=0.42\textwidth]{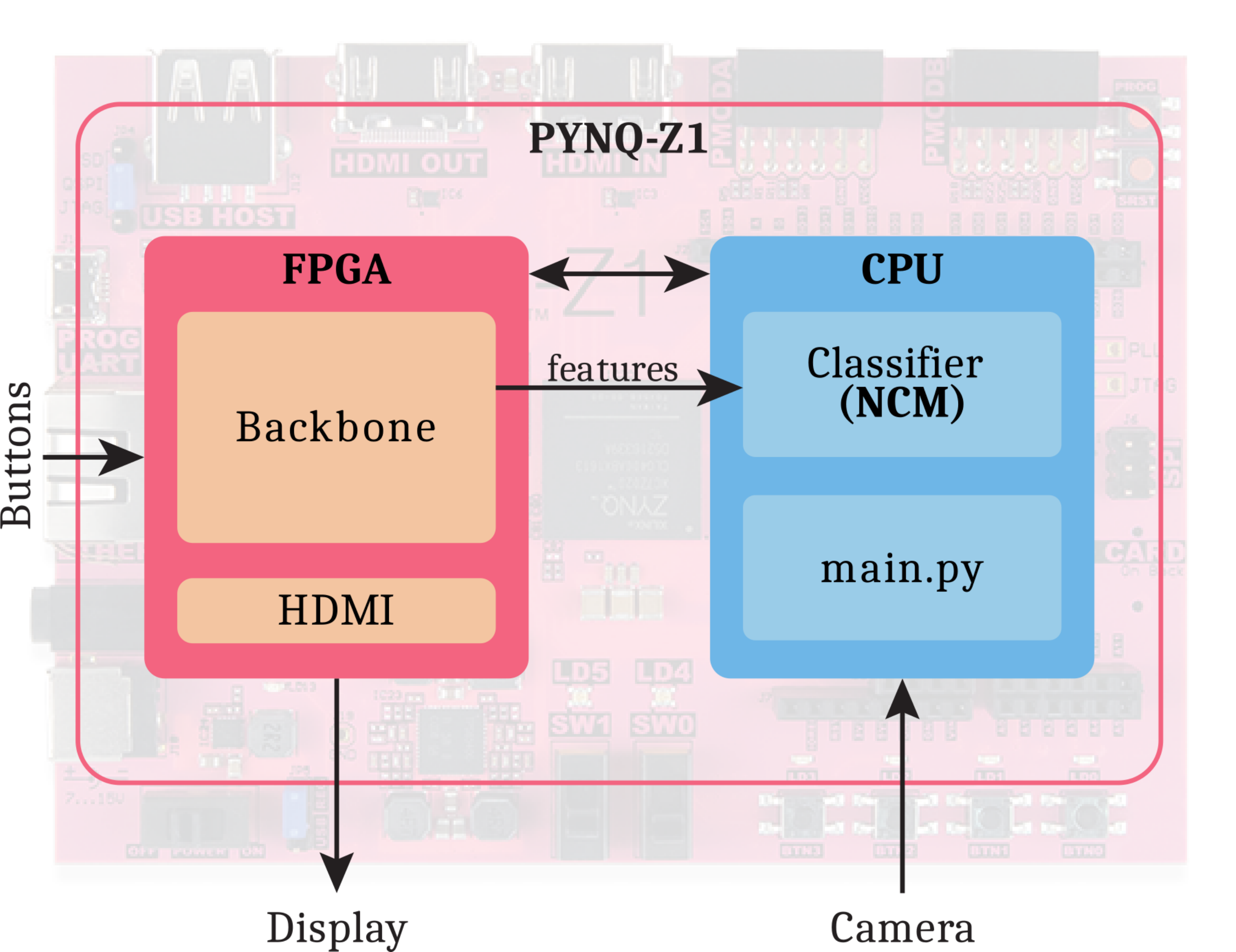}}
\caption{Schematic of the system.}
\label{arch_PYNQ}
\end{figure}

In order to demonstrate how easily this work is applicable in an industrial application context, we created a standalone demonstrator in a compact box. Fig.~\ref{arch_PYNQ} shows a schematic of our demonstrator, it consists of the PYNQ-Z1 board, a 800x540p HDMI screen, a 160x120p camera, and a 10,000mAh battery. It has a 5.75-hour battery life during inference. The demonstration includes on screen indicators for a better user experience. With the entire system and indicators, we achieve an average of 16 FPS during inference. The network is a ResNet-9 with 16 feature maps. The inference runs in the FPGA at 125 MHz, it is implemented using a 16-bit fixed-point format with 8 bits designated for the integer part. The entire system, encompassing the SoC, camera, and screen, operates with a power consumption of 6.2W. On the programmable logic are implemented a Tensil hardware accelerator and an HDMI Xilinx IP which are using most of the FPGA resources. Then, all the software including pre-processing, post-processing, and image classification is executed on the CPU. The demonstrator includes interfaces to camera and buttons to control a live demo.

For the hardware implementation we use the base architecture proposed by Tensil for the PYNQ-Z1 board, increasing only the size of the systolic array from $8\times8$ to $12\times12$, which corresponds to the highest possible value to fit in the FPGA alongside the HDMI controller. In the current version of the pipeline, the NCM classifier is implemented on the CPU side, in a future version we intend to move it to the FPGA.

\section{Results}
\label{sec:results}
\subsection{Design Space Exploration}

\begin{figure}[t]
\centerline{\includegraphics[width=0.45\textwidth]{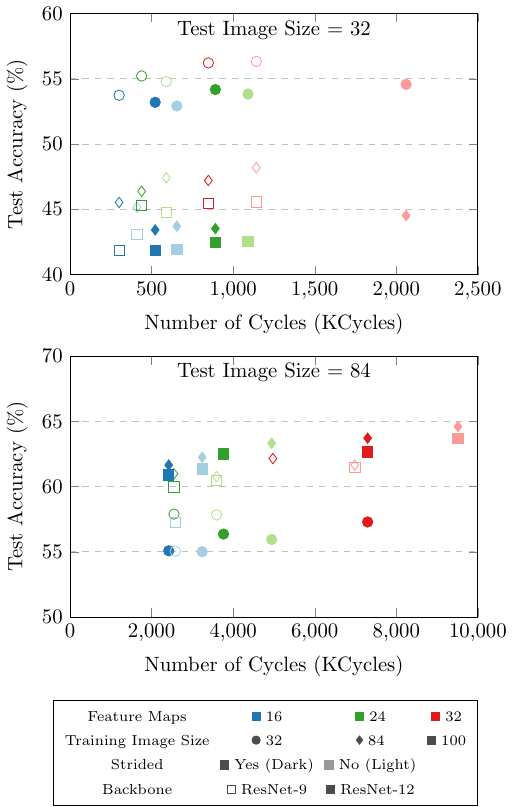}}
\caption{Accuracy and Latency Trade-off: Graphs depict tests on $32\times32$ (top) and $84\times84$ (bottom) images. Different feature maps configurations are denoted by unique colors, while distinct training image sizes are represented by different shapes. We also investigate the impact of strided architectures, differentiated by dark and light colors. Additionally, we vary the backbone architecture from ResNet-9, with empty forms, and ResNet-12, filled forms.}
\label{result_implem}
\end{figure}

The training results are presented in Fig.~\ref{result_implem}. The hyperparameters search space defined in section \ref{sec:backbones} was exhaustively explored. We compiled each network with Tensil to obtain the number of cycles taken by the network's inference. In order to get a smooth video stream (greater than 10 FPS), it is necessary to work with $32\times32$ images. Therefore, we show the results for this resolution alongside the $84\times84$ resolution that is classically chosen for experiments on the MiniImageNet dataset, in a 5-ways, 1-shot setup. The first takeaway is that for the $32\times32$ resolution, ResNets-9 (empty marks) exhibit higher accuracies than the ResNets-12 (full marks), despite their lower number of layers and parameters. We hypothesized that, with $32\times32$ images on a ResNet-12, the dimension of feature maps in the last layer is too small and hardly exploitable by the downstream NCM.

The second takeaway is that for a target test resolution of $32\times32$, the training image sizes should be the same, $32\times32$ (circles). Indeed, the networks trained on larger images $84\times84$ and $100\times100$ are far less accurate. This could be counter-intuitive as training with images resized on $32\times32$ means that some information is lost in the dataset. It is possible that an even better accuracy could be obtained with clever data augmentation or a better generalization error metric~\cite{bendou2023statistical}.

Using convolutions with a stride of 2 in the network allows for a reduction in the number of operations to be executed when compared to using max pooling layers to reduce the dimensions of intermediate representations in the network. This is denoted as \emph{strided} in the Fig~\ref{result_implem}. We verify that the latency is reduced in this case, but also that the accuracy is not impacted by this change, if not increased. Finally, the number of feature maps of the first layer, used as a way to scale the width of the network, allows for a trade-off between latency and accuracy.


In summary, for our specific application, the optimal trade-off lies in the top-left corner, where we can identify configurations with acceptable accuracy and the lowest latency. Consequently, we have selected the strided ResNet-9, trained with $32\times32$ images and 16 feature maps, utilizing $32\times32$ images during inference, empty blue circle on the first graph of Fig.~\ref{result_implem}.




\subsection{Comparison with other hardware implementations}

We set the array size of our systolic array to 12, which corresponds to the maximum possible array size for our setup. The FPGA frequency has been set to 125MHz. Under this configuration, the latency of the backbone inference is 30ms. Few articles have specifically addressed few-shot learning on FPGAs or in embedded systems. An example of few-shot pest recognition on an FPGA has been proposed in~\cite{b3}, reaching 2 frames per second on a PYNQ-Z1. To demonstrate that the hardware resources and latency obtained using Tensil's framework, based on the computational complexity of our backbone, are within the standards of the literature, we conducted a benchmark and present the results in Table~\ref{tab:implem}. We decided to compare with implementations of DNNs proposed for classification on the CIFAR-10 dataset~\cite{krizhevsky2009learning}. Indeed, these are images with a resolution of $32\times32$ pixels, for which the backbone we have chosen (ResNet-9 with 16 feature maps) is highly adaptable, provided that we add a downstream linear layer. We restricted our search to implementations on the same chip as ours, the Zynq-7020 (z7020). From Table~\ref{tab:implem} shows that Tensil's implementation offers comparable latency and accuracy for equivalent resources, validating our backbone. It is important to notice each work implements a different DNN. For this benchmark, we use array size of 12 at 50 MHz.


\begin{table}[]
\centering
\caption{Cifar-10 inference on z7020 FPGA}
\label{tab:implem}
\begin{tabular}{lccccccc}
\hline
\multicolumn{1}{c}{\textbf{Work}} &
  \textbf{\begin{tabular}[c]{@{}c@{}}Prec.\\ {[}bits{]}\end{tabular}} &
  \textbf{LUT} &
  \textbf{\begin{tabular}[c]{@{}c@{}}BRAM \\ {[}36 kb{]}\end{tabular}} &
  \textbf{FF} &
  \textbf{DSP} &
  \textbf{\begin{tabular}[c]{@{}c@{}}Latency \\ {[}ms{]}\end{tabular}} &
  \textbf{\begin{tabular}[c]{@{}c@{}}Acc. \\ {[}\%{]}\end{tabular}} \\ \hline
\cite{borras2022open} hls4ml & 8-12 & 28544 & 42  & 49215 & 4   & 27.3 & 87 \\
\cite{borras2022open} FINN   & 1    & 24502 & 100 & 34354 & 0   & 1.5  & 87 \\
\cite{yang2018fully}           & 1-2  & 23436 & 135 & -     & 53  & 1.1  & 86 \\
\cite{b5}           & 16  & 15200 & 523 & 41     & 167  & 109  & - \\\\
\textbf{Ours}                    & 16   & 15667 & 59  & 9819  & 159 & 35.9 & 92 \\ \hline
\end{tabular}
\end{table}
\section{Conclusion}

In this paper, we propose the first implementation of inductive few-shot learning system on an FPGA SoC, allowing for fast inference and low power consumption.
We propose PEFSL, a fully open-source implementation pipeline that allows for designing a neural network architecture, training and deploying it on an embedded system. 
Our implementation achieves 54\% accuracy on the MiniImageNet dataset for the $32\times32$ resolution in the 1-shot, 5-ways scenario, with a 30ms latency on the PYNQ-Z1 board.



\bibliographystyle{IEEEtran}
\bibliography{IEEEabrv,main.bib}

\end{document}